\RequirePackage{lineno} 
\documentclass[12pt]{article}
\usepackage{graphicx}

\newcommand\pubnumber{DPF2013-269}
\newcommand\pubdate{\today}

\def\tuscaloosa{Department of Physics and Astronomy\\
University of Alabama, Tuscaloosa, AL 35487, USA
}

\def\Title#1{\begin{center} {\Large #1 } \end{center}}
\def\Author#1{\begin{center}{ \sc #1} \end{center}}
\def\Address#1{\begin{center}{ \it #1} \end{center}}

\newcommand\pubblock{\rightline{\begin{tabular}{l} \pubnumber\\
         \pubdate  \end{tabular}}}
\newenvironment{Abstract}{\begin{quotation}  }{\end{quotation}}
\newenvironment{Presented}{\begin{quotation} \begin{center} 
             PRESENTED AT\end{center}\bigskip 
      \begin{center}\begin{large}}{\end{large}\end{center} \end{quotation}}

\def\beq{\begin{equation}}
\def\eeq#1{\label{#1}\end{equation}}
\def\eeqn{\end{equation}}

\def\beqa{\begin{eqnarray}}
\def\eeqa#1{\label{#1}\end{eqnarray}}
\def\eeqan{\end{eqnarray}}

\let\bar=\overbar

\def\Dslash{\not{\hbox{\kern-4pt $D$}}}
\def\dslash{\not{\hbox{\kern-2pt $\del$}}}

\def\msb{{\bar{\ssstyle M \kern -1pt S}}}

\begin{document}
\begin{titlepage}
\pubblock

\vfill
\Title{The Precision IceCube Next Generation Upgrade}
\vfill
\Author{ Dawn Williams}
\Author{For the IceCube, PINGU Collaboration \footnote{For the full IceCube author list see http://icecube.wisc.edu/collaboration/authors/current.}}
\Address{\tuscaloosa}
\vfill
\begin{Abstract}
The IceCube Neutrino Observatory, completed in December 2010 and located at the
geographic South Pole, is the largest neutrino telescope in the world.
IceCube includes the more densely instrumented DeepCore subarray, which
increases IceCube's sensitivity at neutrino energies down to 10 GeV.
DeepCore has recently demonstrated sensitivity to muon neutrino
disappearance from atmospheric neutrino oscillation. A further extension
is under consideration, the Precision IceCube Next Generation Upgrade
(PINGU) which would lower the energy threshold below about 10~GeV. In particular, PINGU would
be sensitive to the effects of the neutrino mass hierarchy, which is one
of the outstanding questions in particle physics. Preliminary
feasibility studies indicate that PINGU can make a high significance
determination of the mass hierarchy within a few years of construction.
\end{Abstract}
\vfill
\begin{Presented}
DPF 2013\\
The Meeting of the American Physical Society\\
Division of Particles and Fields\\
Santa Cruz, California, August 13--17, 2013\\
\end{Presented}
\vfill
\end{titlepage}
\def\thefootnote{\fnsymbol{footnote}}
\setcounter{footnote}{0}

\section{Introduction}

One of the outstanding problems in neutrino physics is the nature of
the neutrino mass hierarchy (NMH), or the sign of $\Delta m_{13}^2$. A
number of experiments at accelerators are in planning for the purposes
of determining the NMH. Megaton Cherenkov neutrino detectors in ice or
water provide another potential path to the NMH measurement. 
\subsection{IceCube DeepCore}
\begin{figure}[htb]
\centering
\includegraphics[height=3.0in]{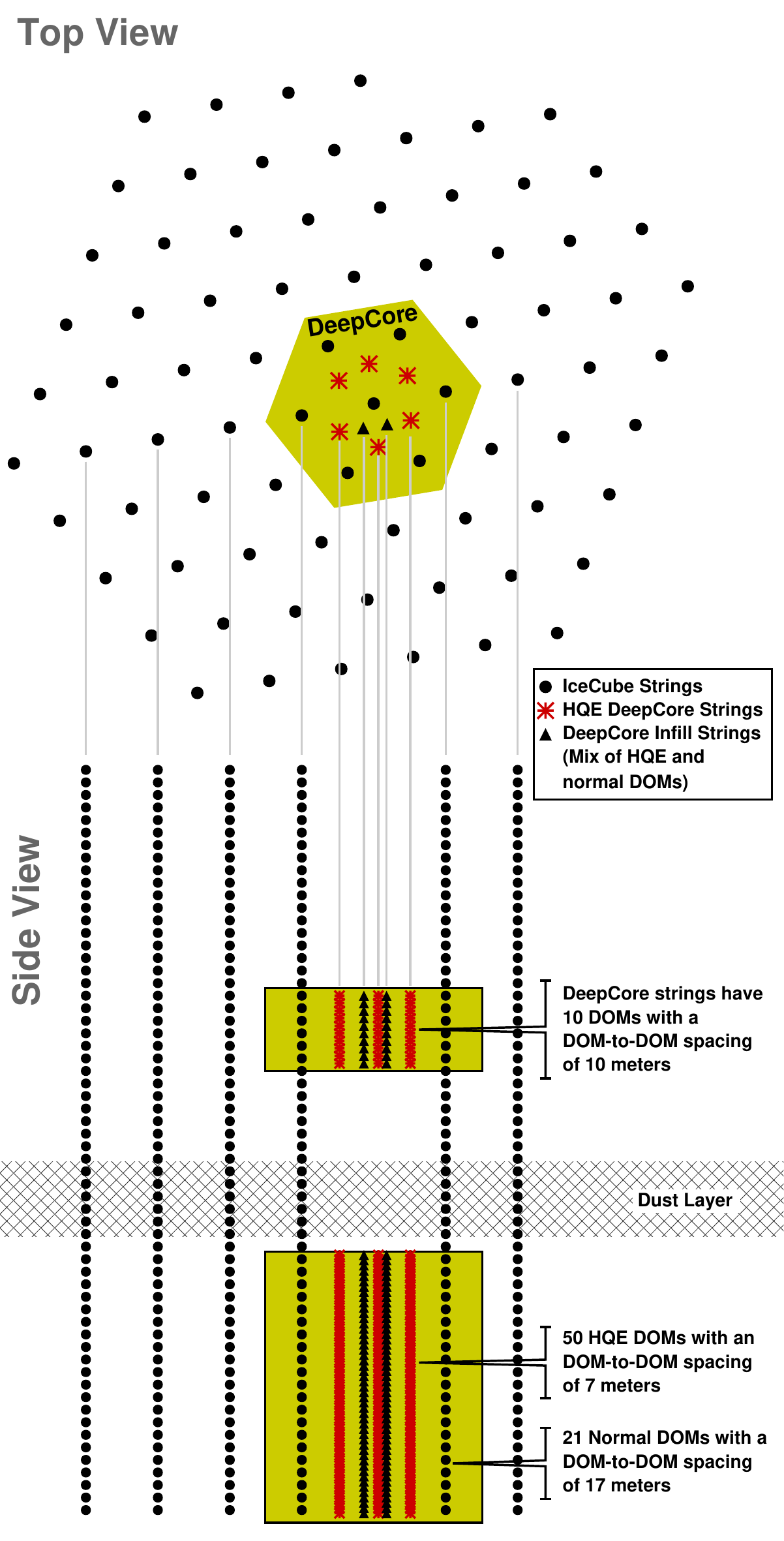}~\hspace{0.5in}~\includegraphics[height=3.0in]{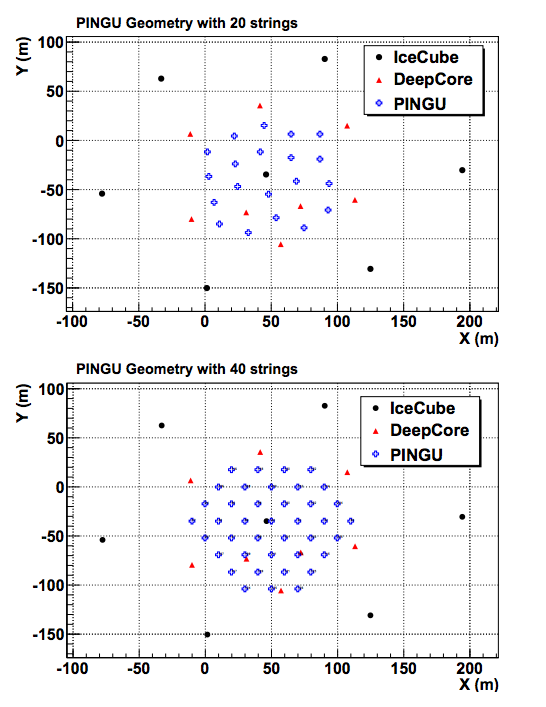}
\caption{Left: IceCube including DeepCore as seen from the top and
  side. Black circles: standard IceCube strings. Red stars and black
  triangles: DeepCore. Right: 20-string and 40-string
  PINGU geometries under study. Black circles: standard IceCube
  strings. Red triangles: DeepCore strings. Blue crosses: PINGU strings.}
\label{fig:pingugeo}
\end{figure}

The IceCube Neutrino Observatory, located at the geographic
South Pole, is the largest neutrino detector in the world, designed to
detect ultra high-energy neutrinos from astrophysical sources such as
active galactic nuclei and gamma ray bursts. IceCube
consists of 86 cables called ``strings'', each instrumented with 60
Digital Optical Modules (DOMs). The DOMs are deployed between 1450~m
and 2450~m deep in the Antarctic ice. Each DOM consists of a glass
pressure vessel containing a 10-inch
photomultiplier tube (PMT), digitizing electronics, and LED flashers for
calibration. IceCube
includes a surface component (IceTop) and a more densely instrumented inner array called DeepCore,
which was designed to improve IceCube's sensitivity to neutrinos at
low energies~\cite{Collaboration:2011ym}. DeepCore strings are located
in the center of IceCube, with smaller string-to-string and DOM-to-DOM
spacing than the surrounding IceCube
strings, as shown in Figure~\ref{fig:pingugeo}. DeepCore DOMs are concentrated below 2000~m depth, where
the ice is the most transparent to light deposited by neutrino
interactions in the ice. Most DeepCore DOMs contain PMTs which have 35\% higher quantum
efficiency than standard IceCube PMTs. The closer spacing and higher
efficiency increase DeepCore's sensitivity to neutrinos below 50~GeV
in energy. DeepCore's physics program
includes indirect detection of dark matter~\cite{Aartsen:2012kia},
galactic supernovae~\cite{Demiroers:2011am}, and
atmospheric neutrino physics~\cite{Aartsen:2013jza}. 

The amount and pattern of light detected by IceCube DOMs is used to
reconstruct the properties of the incident particle: direction,
position, time, and energy. Neutrino event topologies in IceCube fall into two categories:
track-like events from charged current (CC) muon neutrino
interactions, and shower-like events, or cascades, from electron and tau neutrino CC events
and neutral current (NC) events of all flavors. To detect neutrino
events in DeepCore, the surrounding IceCube strings are used to
veto the background from cosmic ray-induced muons which has a $10^6$
times higher rate at trigger level than atmospheric
neutrinos. 
\subsection{Atmospheric Neutrino Physics in DeepCore}
DeepCore recently reported the first
statistically significant observation of muon neutrino disappearance at
energies above 20~GeV~\cite{Aartsen:2013jza}. The disappearance
signature in DeepCore is a
deficit of muon neutrino tracks at upgoing zenith angles due to  a
minimum in the survival probability of Earth-crossing muon neutrinos at 25~GeV. The fitted oscillation parameters from
the muon neutrino deficit are consistent with the world average values
for $\Delta m^{2}_{23}$ and sin$^{2}(2\theta_{23})$. IceCube also successfully isolated low energy
cascades from atmospheric electron neutrinos and NC interactions in
DeepCore, reporting the first measurement of the atmospheric electron
neutrino flux between 80~GeV and 6~TeV~\cite{Aartsen:2012uu}.

\section{PINGU}

The Precision IceCube Next Generation Upgrade (PINGU) is a proposed
infill extension of IceCube. PINGU would add strings with 60-100
additional DOMs each inside of the
DeepCore volume. Detector geometries with 20 and 40 additional
strings are under study, shown in
Figure~\ref{fig:pingugeo}. These strings can be deployed within 2-3
years for a relatively modest cost, based on the construction experience of IceCube~\cite{pingu_snomass}. The basic
design of PINGU is that of IceCube, with several improvements planned including
a simplified digitization scheme, an upgraded system of calibrated
{\it in situ} light sources, and degassing of the drill water during
the drilling process in order to mitigate the formation of bubbles in
the PINGU holes. PINGU would
increase IceCube's effective volume for neutrinos at energies below
20~GeV.


\subsection{Neutrino Mass Hierarchy with PINGU}
\begin{figure}[htb]
\centering
\includegraphics[height=2.3in]{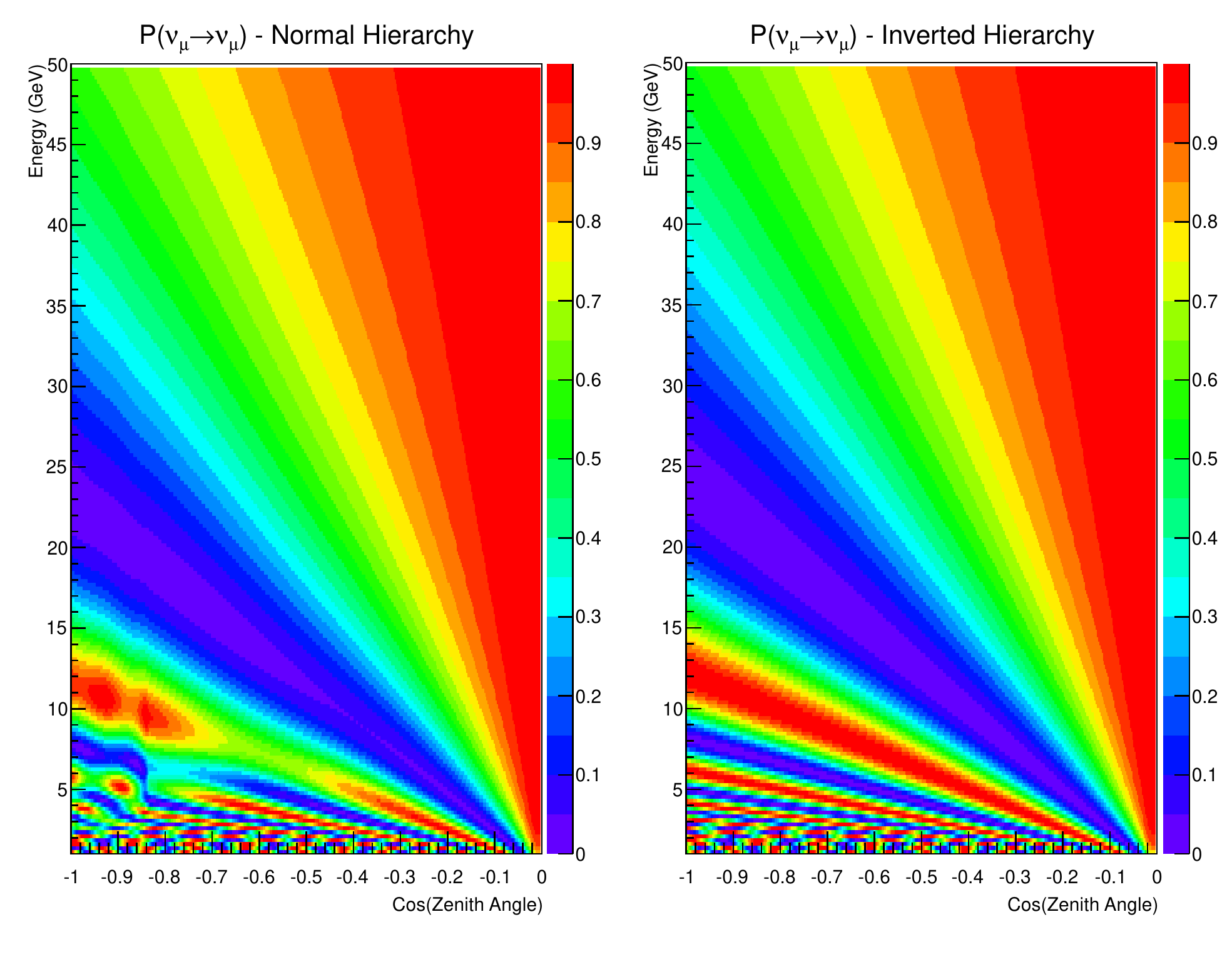}~\includegraphics[height=2.3in]{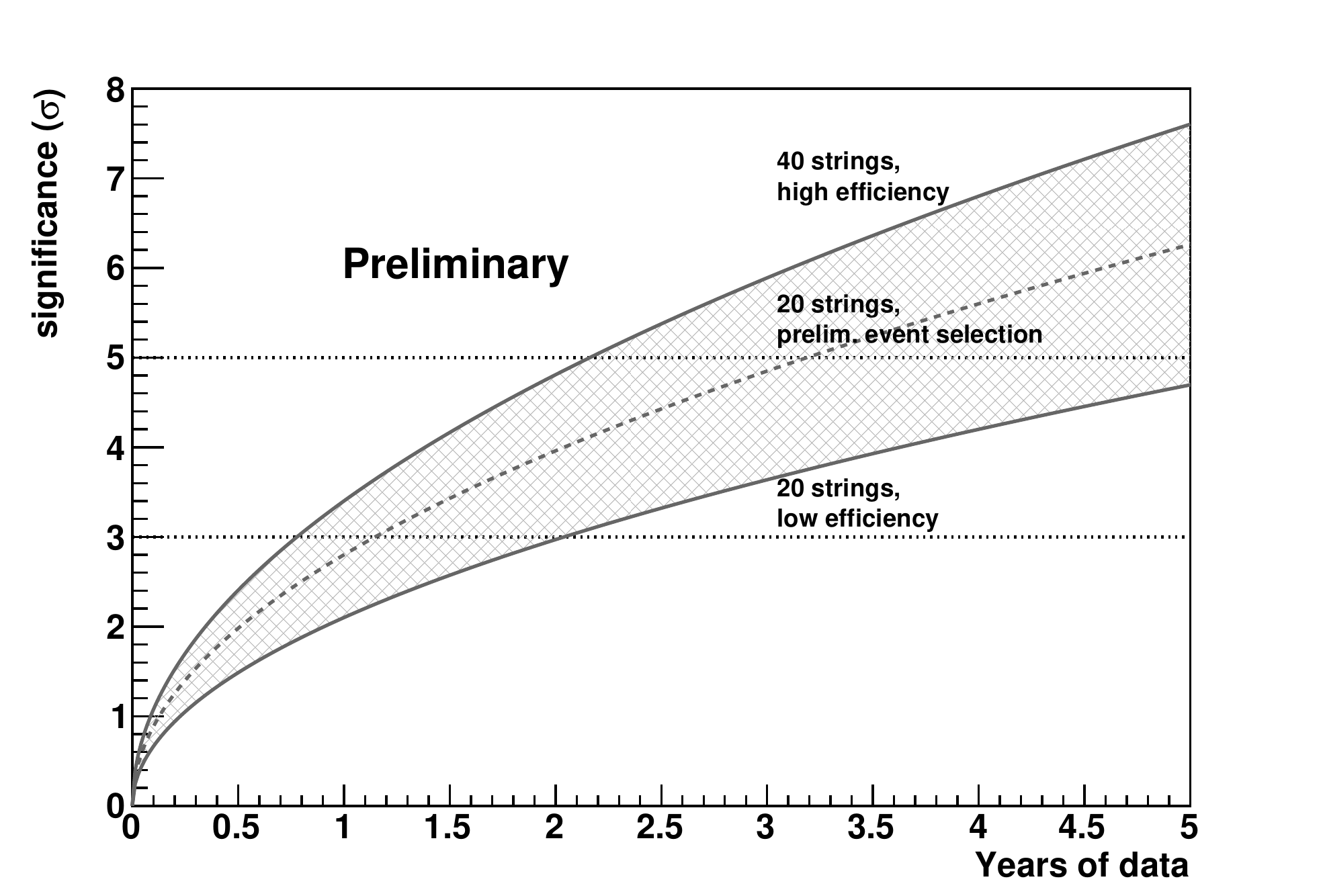}
\caption{Left: muon neutrino survival probability as a function of
  neutrino energy and the cosine of the zenith angle (cos($\theta_Z$) =
  -1 for upgoing
  neutrinos) in the normal and
inverted hierarchies, with enhanced oscillation probabilities due to
neutrino propagation through the Earth visible in the normal
hierarchy. For antineutrinos, enhancements would occur in the inverted
hierarchy. Right: significance of NMH measurement in PINGU as a
function of time. The band encompasses the results of three
independent studies.}
\label{fig:oscil}
\end{figure}

Neutrinos oscillating in the Earth undergo the MSW
effect~\cite{msw}, which modifies the oscillation probability
of neutrinos propagating through matter. 
The oscillation probability of (anti-)neutrinos is enhanced in the
normal (inverted) mass hierarchy. Neutrino oscillation probabilities also undergo parametric
enhancement at density boundaries, such as the boundary between the
core and the mantle~\cite{parametric}. For Earth-crossing neutrinos,
the enhancement is strongest below 10~GeV. The relatively large
value of $\theta_{13}$~\cite{pdgreview} makes the measurement of the
NMH feasible in PINGU~\cite{Mena:2008rh, Winter:2013ema, akhmedov}. The muon neutrino
survival probability in PINGU as a function of zenith angle and
energy, for both normal and inverted hierarchies, is shown in Figure~\ref{fig:oscil}. Although PINGU cannot distinguish between neutrinos and
antineutrinos, differences in the cross section and fluxes for
neutrinos vs. antineutrinos produce a measurable difference in the
muon neutrino flux as a function of 
energy and zenith angle for the normal and inverted hierarchies.

The distinguishability metric defined in ~\cite{akhmedov} may be used to quantify the observable
difference between the normal and inverted hierarchies:

\begin{equation} \label{eq:distinguish}
S_{tot}=\sqrt{\frac{(N_{i,j}^{IH}-N_{i,j}^{NH})^2}{N_{i,j}^{IH}}}
\end{equation}

where $N_{i,j}$ is the number of muon neutrino events in the
$i,j$th bin in neutrino energy and cosine of zenith angle (cos($\theta_Z$)). The
distinguishability metric indicates which regions in the
$E_{\nu}-cos(\theta_Z)$ plane have the most sensitivity to the NMH. The
significance of a NMH measurement in PINGU is computed taking
reconstruction efficiency and
systematic uncertainties into account.

\begin{figure}[htb]
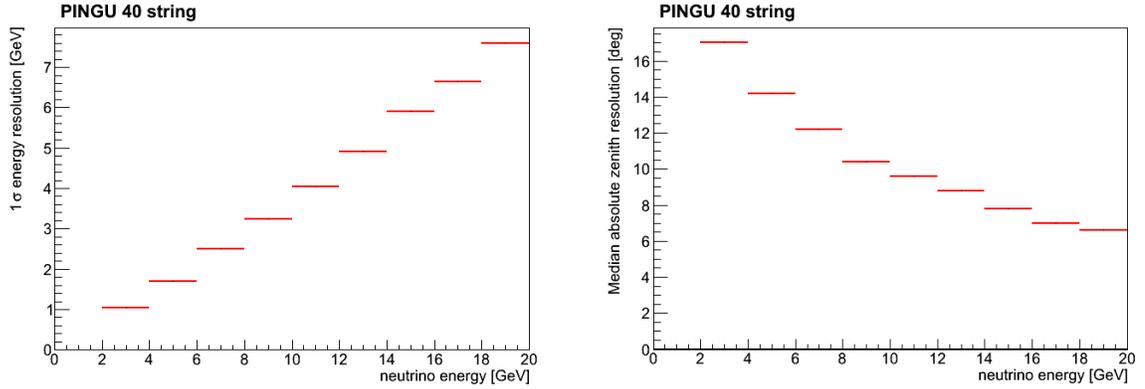

\centering
\includegraphics[height=2.1in]{monopod}~\includegraphics[height=2.1in]{santa}
\caption{Left: energy resolution as a function of energy in the PINGU
  40-string geometry, using a maximum likelihood reconstruction algorithm. Right: angular
  resolution as a function of energy using an algorithm which
  detects unscattered light from muons. }
\label{fig:reco}
\end{figure}

To determine the energy and angular resolution, low-energy reconstruction algorithms developed for DeepCore data
analysis have been applied to simulated PINGU events. Results from
these algorithms with the baseline 40-string
geometry are shown in Figure~\ref{fig:reco}. A direction reconstruction
algorithm using unscattered light from muons yields a median angular
resolution of $15^{\circ}$ at 5~GeV, improving to $8^{\circ}$ at
20~GeV. A maximum likelihood energy reconstruction yields an energy
resolution of 0.7 GeV +
0.2×$E_{\nu}$(GeV). Flavor identification algorithms
are under study which identify the muon track by its speed, which is
near the speed of light in vacuum, as opposed to light from
cascade-like events which travels at the speed of light in ice. 

Three independent analyses have estimated the sensitivity of PINGU to
the NMH. Systematic effects used in these studies
include: $\pm2\sigma$ variation on the world average values of
$\theta_{23}$, $\theta_{13}$,$\Delta m^{2}_{atm}$ and $\delta_{CP}$;
30\% uncertainty in the effective volume; $\pm 0.05$ in the
atmospheric neutrino spectral index; 10\% error in the energy and
direction reconstruction; and 10\% error in the energy and angular
resolution. The effects of nonzero $\delta_{CP}$ and uncertainties in $\theta_{12}$ and $\delta
m_{12}^2$ are under investigation, but are expected to be
small~\cite{akhmedov}. These studies indicate that the wide range of
energies and baselines available to PINGU allows control of systematic errors.

The sensitivity as a function of detector livetime, displayed
as a band encompassing the range of results from the various sensitivity studies, is shown in
Figure~\ref{fig:oscil}.  We estimate that PINGU will have roughly 3$\sigma$
sensitivity to the neutrino mass hierarchy within 2 years of
completing construction~\cite{pingu_snomass}. 


\section{Summary}
PINGU has the potential to measure the neutrino mass hierarchy on a
relatively short timescale for a relatively modest cost. This measurement would be complementary to
long baseline accelerator neutrino experiments in planning or under
construction. Further studies of systematic effects are under investigation
and a letter of intent is in preparation, with a full proposal to
follow. 


\end{document}